\documentclass[aip, apl, superscriptaddress, preprint]{revtex4-1}

\usepackage{graphicx}
\usepackage{bm}
\usepackage{color, soul}
\usepackage{amsmath}
\usepackage{epstopdf}

\begin{document}

\title{Spin chirality fluctuation in two-dimensional ferromagnets with perpendicular anisotropy}

\date{\today}

\author{Wenbo Wang$^\ast$}
\affiliation{Department of Physics and Astronomy, Rutgers University, Piscataway, New Jersey 08854, USA}
\author{Matthew W. Daniels$^\ast$}
\affiliation{Institute for Research in Electronics and Applied Physics, University of Maryland}
\affiliation{Department of Physics, Carnegie Mellon University, Pittsburgh, Pennsylvania 15213, USA}
\author{Zhaoliang Liao$^\ast$}
、\affiliation{National Synchrotron Radiation Laboratory, University of Science and Technology of China, Hefei, 230026 Anhui, People's Republic of China}
\affiliation{MESA+ Institute for Nanotechnology, University of Twente, 7500 AE Enschede, the Netherlands}
\author{Yifan Zhao}
\affiliation{Department of Physics, Pennsylvania State University, University Park, Pennsylvania 16802, USA}
\author{Jun Wang}
\affiliation{MESA+ Institute for Nanotechnology, University of Twente, 7500 AE Enschede, the Netherlands}
\author{Gertjan Koster}
\affiliation{MESA+ Institute for Nanotechnology, University of Twente, 7500 AE Enschede, the Netherlands}
\author{Guus Rijnders}
\affiliation{MESA+ Institute for Nanotechnology, University of Twente, 7500 AE Enschede, the Netherlands}
\author{Cui-Zu Chang}
\affiliation{Department of Physics, Pennsylvania State University, University Park, Pennsylvania 16802, USA}
\author{Di Xiao}
\affiliation{Department of Physics, Carnegie Mellon University, Pittsburgh, Pennsylvania 15213, USA}
\author{Weida Wu}
\email[Corresponding author: ]{wdwu@physics.rutgers.edu}
\affiliation{Department of Physics and Astronomy, Rutgers University, Piscataway, New Jersey 08854, USA}


\begin{abstract}
Non-coplanar spin textures with scalar spin chirality can generate effective magnetic field that deflects the motion of charge carriers, resulting in topological Hall effect (THE), a powerful probe of the ground state and low-energy excitations of correlated systems~\cite{wen89,taguchi01,neubauer09}. However, spin chirality fluctuation in two-dimensional ferromagnets with perpendicular anisotropy has not been considered in prior studies. Herein, we report direct evidence of universal spin chirality fluctuation by probing the THE above the  transition temperatures in two different ferromagnetic ultra-thin films, SrRuO$_3$ and V doped Sb$_2$Te$_3$.  The temperature, magnetic field, thickness, and carrier type dependences of the THE signal, along with our Monte-Carlo simulations, unambiguously demonstrate that the spin chirality fluctuation is a universal phenomenon in two-dimensional ferromagnets with perpendicular anisotropy.  Our discovery opens a new paradigm of exploring the spin chirality with topological Hall transport in two-dimensional magnets and beyond~\cite{Bonilla2018,Fei2018,Deng2018,machida10}.  
\end{abstract}

\maketitle

Understanding quantum transport of electrons in magnets is a fundamental issue in strongly correlated systems and spintronics~\cite{wen89,taguchi01,neubauer09}.  If the magnetic moments form a non-coplanar spin texture as in many frustrated and/or chiral magnets, an electron traveling through the system will experience an effective magnetic field originated from the real-space Berry phase as it hops along a loop of three neighboring magnetic moments (called a triad)~\cite{taguchi01}.  Thus, this effective field is proportional to the scalar spin chirality of the triad, and gives rise to a transverse response known as the topological Hall effect (THE). The THE is distinguished from both the ordinary Hall effect (OHE), which requires the application of an external magnetic field, and the anomalous Hall effect (AHE) found in ferromagnets with a uniform magnetization, in which the dominant contribution comes from the Berry phase in the momentum space {in moderately conducting samples}~\cite{Nagaosa2010}.  Because of its close relation to non-coplanar spin textures, the THE is a powerful probe to detect exotic phases in magnetic systems.  Indeed, the THE has been instrumental in the electric detection of skyrmion phase in chiral magnets and heterostructures~\cite{neubauer09,zang2011,Kanazawa2011,Huang2012}. 

While the THE is l{often} interpreted as a signature of static spin textures with spin chirality {in two-dimensional (2D) magnetic thin films}, it has been proposed that thermal fluctuations of topological excitations could also result in a significant Hall effect in three-dimensional magnets such as manganites~\cite{matl98,ye99,Chun2000}.
However, the temperature dependence of the observed Hall resistivity cannot exclude the conventional mechanisms~\cite{taguchi01}. 
Herein, we report direct evidence of universal spin chirality fluctuation in  {2D} ferromagnets with perpendicular anisotropy by probing the THE.   Substantial THE signal was observed above the ferromagnetic transition temperatures $T_\mathrm{c}$'s in two completely different itinerant ferromagnets, SrRuO$_3$ (SRO) and  V {(5\%)} doped Sb$_2$Te$_3$ (VST) thin films in the 2D limit.  Remarkably, the observed THE persists well into the paramagnetic phase, clearly demonstrating a thermal-fluctuation origin. 
SRO is a metallic ferromagnet with $n$-type carriers~\cite{Koster2012}, while VST is a magnetically doped topological insulator with $p$-type carriers~\cite{Chang2015}.    
The temperature ($T$), magnetic field ($H$), thickness ($t$), and carrier type dependence of the THE are in excellent agreement with our Monte Carlo (MC) simulations, further corroborating the universality of the spin chirality fluctuation in 2D ferromagnets with perpendicular anisotropy.  Given the ubiquitous existence of chiral spin order and fluctuation in correlated systems, our finding  opens the door to exploring the spin chirality with topological Hall transport in 2D ferromagnets~\cite{Bonilla2018,Fei2018,Deng2018} or quantum spin liquids~\cite{wen89,machida10}.

The SRO thin films were grown on STO (001) substrates using pulsed laser deposition, while the VST films were grown on STO (111) substrates with molecular beam epitaxy. We first focus on SRO films to illustrate the discovery of the chiral fluctuation driven THE.  A cartoon schematic of the SRO thin film structure and device configuration is shown in Fig.~1\textbf{a}. Electric contacts were fabricated with wire bonding and silver paint.  The STO capping layer helps to enhance the ferromagnetic ordering in the ultra-thin limit~\cite{Thomas2017}. Metallic behavior was observed in all STO capped SRO films with $t\geq 3$~u.c., while the 2~u.c.\ one is a non-magnetic insulator (See Supplementary Fig.~S2 for transport data).
The Curie temperature $T_\mathrm{c}$ is characterized by the peak anomaly of the slope of the longitudinal resistance ($dR_{xx}/dT$)~\cite{Klein1996,Shen2015}. Thinner SRO films are more resistive with lower $T_\mathrm{c}$, which is consistent with the effect of reduced dimensionality~\cite{Koster2012}.  

In general, there are three contributions to the Hall resistivity in a magnetic metal: the OHE proportional to $H$, the conventional AHE proportional to the magnetization $M$, and the THE due to the real-space Berry phase~\cite{neubauer09,Kanazawa2011,Huang2012}.  Therefore we can express the Hall resistivity as
\begin{equation}
\rho_{yx}(H)=R_{0}H+R_{S}M+\rho_{H}^\mathrm{T}.
\label{e:rho}
\end{equation}
{Here we will focus on the anomalous part: $\tilde\rho_{yx} \equiv \rho_{yx} - R_0 H$ by removing the OHE contribution.}  Well below $T_\mathrm{c}$, a hysteresis loop of $\tilde\rho_{yx}$ was observed for all SRO (and VST) films (See Supplementary information Section I-K).   This loop has the same shape as the magnetization hysteresis loop, indicating that below $T_\mathrm{c}$ the main contribution to $\tilde\rho_{yx}$ comes from the conventional AHE.  This is similar to previous reports of $\tilde\rho_{yx}$ on magnetic thin films with perpendicular anisotropy~\cite{yasuda16,matsuno16,liu17,Chang2015,Zhao2018}.

\begin{figure*}[htp]
\includegraphics[width=.9\textwidth]{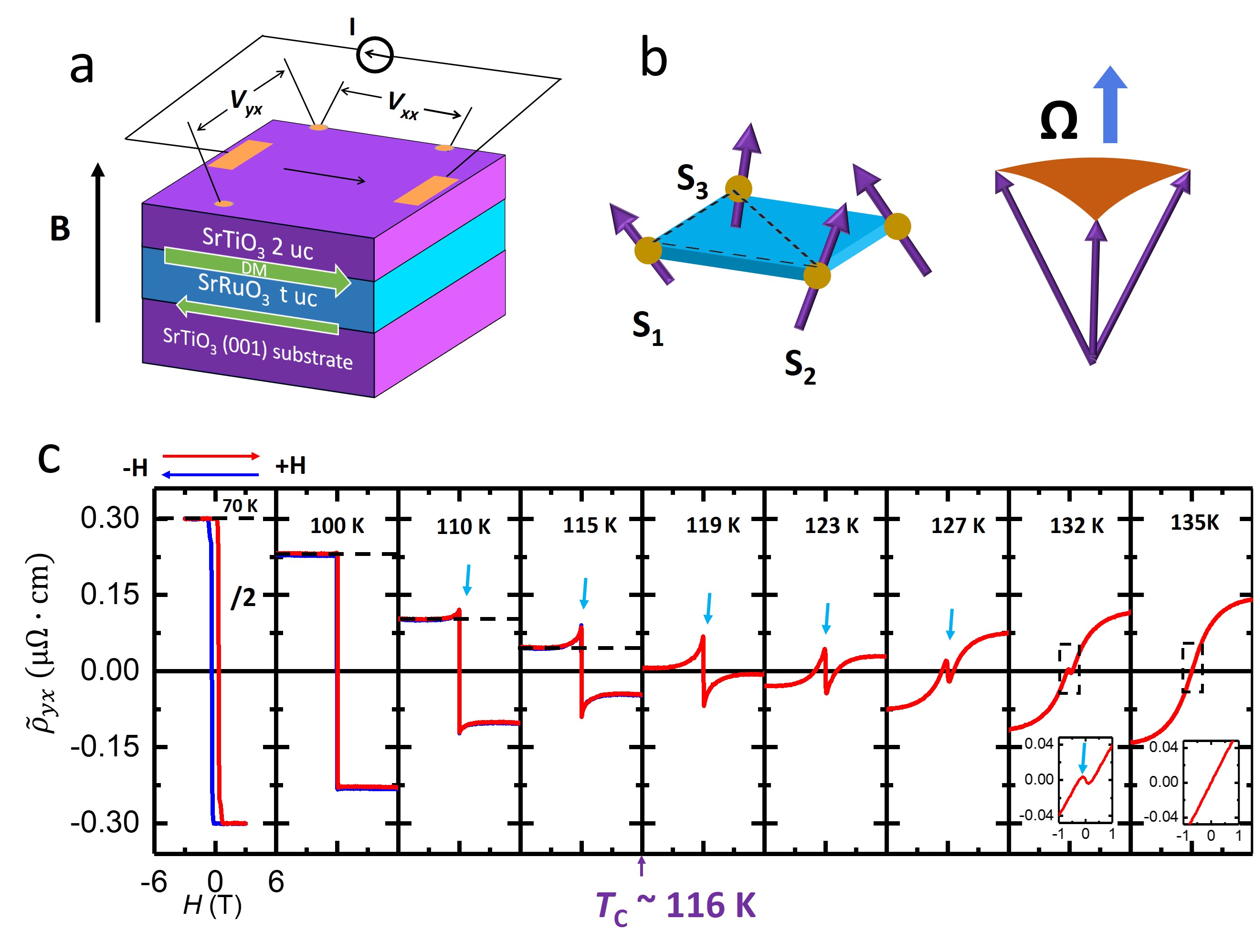}
\caption{\textbf{$\tilde{\rho}_{yx}(H)$ data of STO capped 6 u.c.\ SRO film.} $\textbf{a}$, A schematic of the (STO)$_2$/(SRO)$_t$ bilayer grown on STO (001) and leads for transport measurements. \textbf{b}, a snapshot of non-coplaner spins on a square lattice. The three non-coplanar neighbouring spins subtend a solid angle $\Omega$, resulting in effective magnetic field. \textbf{c}, the Hall resistivity $\tilde{\rho}_{yx}$ (with OHE subtracted) as a function of magnetic field $H$ at various temperatures from 100\,K to 135\,K. Blue (red) curves were taken as the magnetic field from $-$6\, to 6\,T (6\, to $-$6\,T). The cyan arrows indicate the THE humps emerge above 100\,K. The insets show zoomed-in curves of 132 and 135\,K data around zero field, suggesting THE peaks is absent above 135\,K. 
\label{fig1}}
\end{figure*}

Surprisingly, at higher temperatures a sharp anomaly in $\tilde\rho_{yx}(H)$ develops around $T_\mathrm{c}$.  Figure~\ref{fig1}\textbf{c} shows $\tilde\rho_{yx}(H)$ of the STO capped SRO film (6~u.c.).  Below $T_\mathrm{c}\approx116$~K, $\tilde\rho_{yx}(H)$ shows a pronounced square shape, indicating a robust ferromagnetic ordering with a strong uniaxial anisotropy~\cite{Koster2012}.  As $T$ approaches $T_\mathrm{c}$, a prominent antisymmetric peak near zero field emerges.  This emergent feature indicates an additional contribution to the Hall signal which we attribute to the THE (see discussion below).  In particular, the AHE changes its sign around 119~K  as evidenced from the high-field values of $\tilde\rho_{yx}(H)$. This is due to the strong energy dependence of the AHE near the Fermi energy~\cite{Fang2003,matsuno16}. Therefore $\tilde\rho_{yx}(H)$ data at 119~K is dominated by the THE signal (See Supplementary information for the complete dataset).  Clearly, the THE data at 119\,K is different from the conventional AHE, showing a sharp antisymmetric peak near zero field followed by a smooth suppression at high fields.  Remarkably, this antisymmetric peak feature becomes strongest slightly above $T_\mathrm{c}$, then gradually decreases and disappears around 132\,K, well into the paramagnetic phase. {Note that the THE feature is clearly visible in the raw Hall data. (see Supplementary information Section C)} The persistence of THE in the paramagnetic phase above $T_\mathrm{c}$ clearly demonstrates a thermally driven spin chirality fluctuation mechanism~\cite{Hou2017,Bottcher2017}, which is distinct from the THE due to skyrmion phases emerging below $T_\mathrm{c}$ in previous studies~\cite{yasuda16,matsuno16,liu17,Zhao2018}.   

\begin{figure*}
\includegraphics[width=\textwidth]{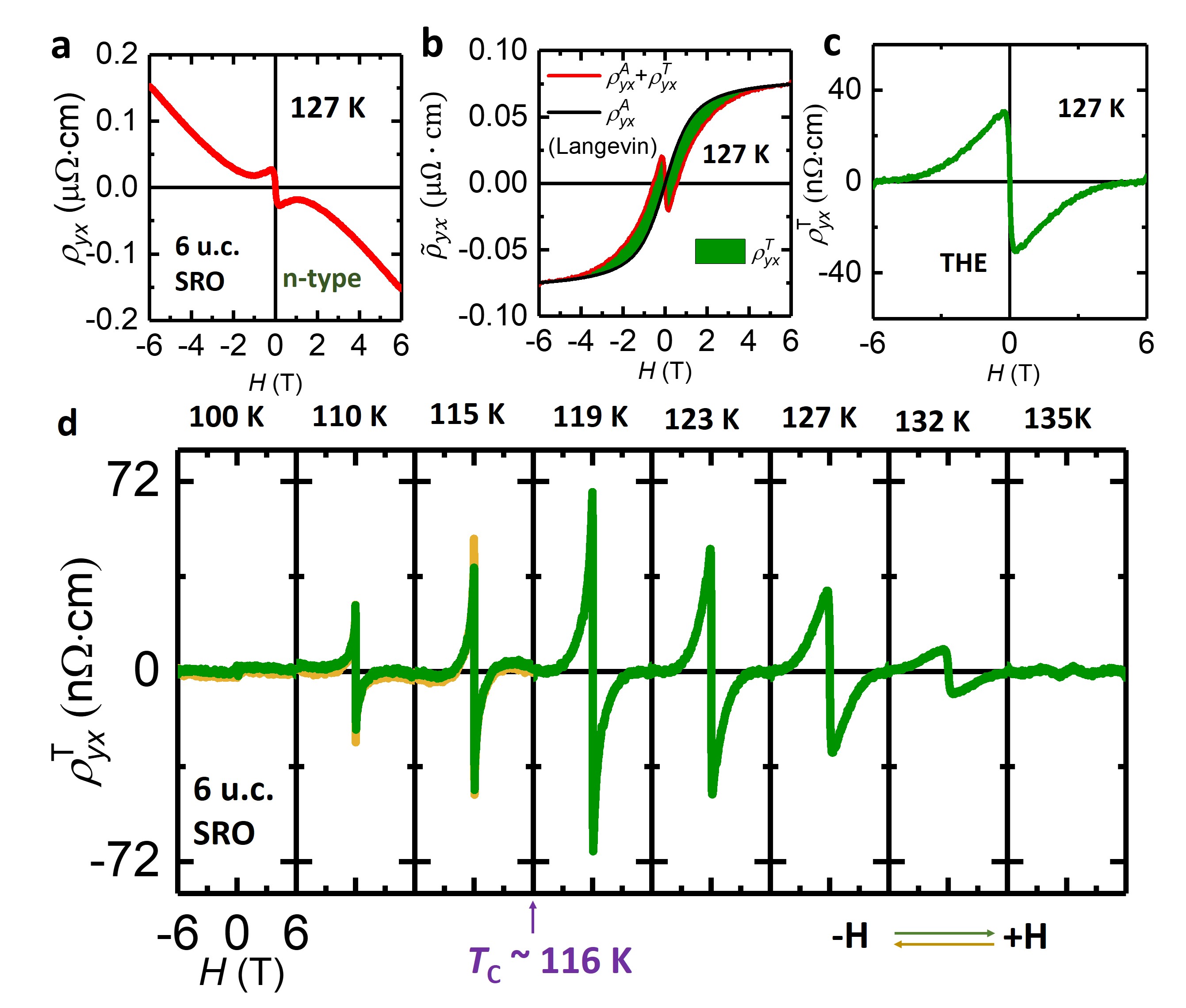}
\caption{\textbf{ THE signal \bm{$\rho^\mathrm{T}_{yx}$} of STO capped 6 u.c.\ SRO film. } 
\textbf{a}, $\rho_{yx}$ of SRO film (6 u.c.) at 127\,K. The negative slope at high field indicate $n$-type carriers.
\textbf{b}, $\tilde\rho_{yx}$ (black) of SRO film at 127\,K was fitted with a Langevin function (red). The green area is the THE signal, which is plotted in 
\textbf{c}. 
\textbf{d}, $\rho^\mathrm{T}_{yx}$ of SRO film as a function of magnetic field $H$ at various temperature from 100\,K to 135\,K.
\label{fig2}}
\end{figure*}

To extract the $T$ evolution of the THE signal, the AHE contribution ($\propto M$) at each $T$ needs to be properly removed. It is technically challenging to directly measure the uniform out-of-plane magnetization of thin films in the 2D limit because of diminishing stray field due to the demagnetization factor~\cite{Hellwig2007}.  For SRO films, we utilize the fact that for the paramagnetic phase at $T> T_\mathrm{c}$, the $M(H)$ curves can be described by a Langevin function in the large spin limit.  Since the the antisymmetric peak feature becomes negligible for $T\ge 135$\,K, these high temperature AHE data are used for the Langevin fitting.  
For $T<T_\mathrm{c}$, a step function is used to approximate the AHE contribution.  Figure~2 shows an example of AHE background subtraction.   The raw data $\rho_{yx}(H)$ at 127~K is shown in Fig.\,\ref{fig2}\textbf{a}. The negative slope at high field indicates $n$-type charge carriers~\cite{Koster2012}. Fig.~\ref{fig2}\textbf{b} shows  $\tilde\rho_{yx}(H)$ after removing the OHE and the Langevin fitting of the AHE background. The difference (green area) between them is the THE signal $\rho^\mathrm{T}_{yx}$, which is plotted in Fig.~\ref{fig2}\textbf{c}.
Using this procedure, we extracted $\rho^\mathrm{T}_{yx}$ of the 6~u.c.\ SRO film at various $T$ as shown in Fig.\,\ref{fig2}\textbf{d}. The antisymmetric THE peaks are visible from 110\,K to 132\,K, and is most pronounced around $T_\mathrm{c}$, which clearly demonstrates that the observed THE originates from thermal fluctuation of spin chirality.

\begin{figure}
\includegraphics[width=\columnwidth]{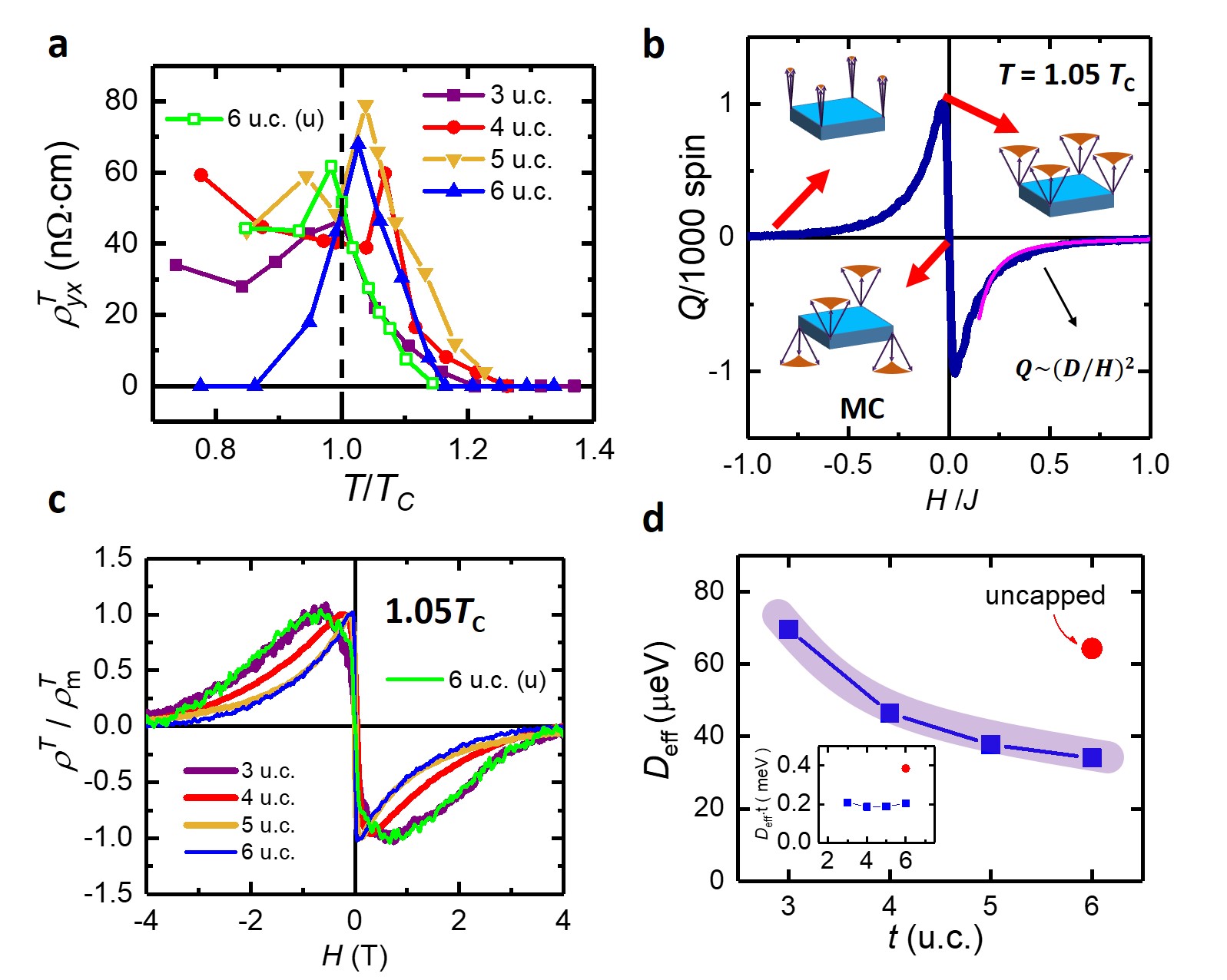}
\caption{\textbf{ Thickness dependence of the THE ($\rho^\mathrm{T}_{yx}$).} \textbf{a}, Maximum values of $\rho^\mathrm{T}_{yx}$ as a function of temperature $T$ of SRO films with thickness $t =3\sim6$\,u.c.\ capped with STO  and the 6 u.c.\ uncapped SRO film.  \textbf{b}, MC simulations ($K=0.5\,J$ and $D=0.25\,J < D_c$) of topological charges $Q$ at $T=1.05T_\mathrm{c}$ shows an antisymmetric profile similar to the experimental data of THE. The high field tail can be described by $(D/H)^2$ (the magenta curve). \textbf{c}, antisymmetric $H$ dependence of normalized $\rho^\mathrm{T}_{yx}$ at the reduced temperature 1.05$T_\mathrm{c}$. \textbf{d}, {$t$ dependence of effective DMI ($D_\mathrm{eff}$). The $D_\mathrm{eff}$ was extracted from fitting the $H$ dependence of $\rho^\mathrm{T}_{yx}$ shown in \textbf{c}. The inset shows the product of $D_\mathrm{eff}\cdot t$, which is approximately a constant, suggesting an interface origin. } 
\label{fig3}} 
\end{figure}

Similar  $T$-$H$ dependence of THE was also observed in thinner SRO films with $t=3,4,5$, corroborating the ubiquitous spin chirality fluctuation induced THE (See Section {I and J} in Supplementary information). {Furthermore, the THE is absent in thicker SRO films ($t>7$), supporting the 2D nature of the chiral fluctuation due to the confinement effect. }
Figure~\ref{fig3}\textbf{a} shows the maximum values of $\rho^\mathrm{T}_{yx}$ as a function of reduced temperature $T/T_\mathrm{c}$. Clearly, the THE signal peaks around $T_\mathrm{c}$, then diminishes approximately 20\% above $T_\mathrm{c}$. The non-monotonic thickness dependence of the THE magnitudes might come from the combined effects of the effective exchange ($J$) and other parameters (\textit{e.g.} anisotropy, magnetic disorders).  The qualitative $T$-$H$ dependences of the THE signal are consistent with the prior theoretical studies of thermal fluctuation in 2D chiral magnets~\cite{Hou2017,Bottcher2017}. However, spin chirality fluctuation in 2D ferromagnets with perpendicular anisotropy has not been addressed.   

To understand the emergent spin chirality fluctuation in 2D {ferromagnets with perpendicular anisotropy}, we carried out Monte Carlo (MC) simulations with the following  Hamiltonian (See Supplemental Materials for more details): 
\begin{equation}
H= \sum_{\langle ij\rangle}\left[-J~(\bm{S}_i\cdot \bm{S}_j)+\bm{D}_{ij}\cdot(\bm{S}_i\times \bm{S}_j)\right]\nonumber\\
-K\sum_{i}(S_i^z)^2 - B_z \sum_i S_i^z \;,
\label{e:Hamitonian}
\end{equation}
where $\bm{S}_i = S\bm{n}_i$ is the spin on the $i$th lattice site. The first term ($J>0$) describes ferromagnetic Heisenberg exchange coupling. The second term describes the chiral interaction, namely the Dzyaloshinsky-Moriya interaction (DMI)~\cite{Dzyaloshinsky1958, Moriya1960}, arising from the inversion symmetry breaking (due to interfaces, \textit{e.g.}).
The uniaxial anisotropy $K$ is included in the third term. With $D<D_c\equiv2\sqrt{2JK}/\pi$, the ground state is a ferromagnet with uniform out-of-plane (OOP) magnetization despite of the presence of chiral interaction (DMI)~\cite{Rohart2013}. The last term is the Zeeman energy due to the external magnetic field $B_z$.

It appears that the DMI is forbidden by the symmetric structure (STO/SRO/STO) of our films.  However, it has been shown that the STO film grown on SRO is slightly different from the STO substrate~\cite{Hyun2001}. This slight difference breaks the OOP inversion symmetry, allowing a small but nonzero chiral interaction (DMI).  Thus,  non-zero spin chirality emerges when the ferromagnetic order is ``melted'' by thermal fluctuations around $T_\mathrm{c}$. The emergent spin chirality is characterized by the topological charge $Q$ defined as:
\begin{equation}
Q = \frac{1}{4\pi}\int d^2\bm r\, \bm n \cdot (\partial_x \bm n \times \partial_y \bm n) \;,
\end{equation}
where $\bm n$ is a unit vector describing the local spin direction.  On a lattice, the integral is replaced by the sum of the solid angle $\Omega$ subtended by three neighboring spins (a triad) as shown in Fig.~1\textbf{b}~\cite{Berg1981}.  As shown in Fig.~\ref{fig3}\textbf{b}, our MC simulations demonstrate that the topological charge density $Q(H)$ is nonzero around $T_\mathrm{c}$, showing qualitatively the same antisymmetric peak structure as that of the observed THE (see Supplementary information for details). The excellent agreement clearly demonstrates that the observed THE originates from the effective magnetic field generated by the thermal-fluctuation driven spin chirality.

\begin{figure}[htp]
\includegraphics[width=\columnwidth]{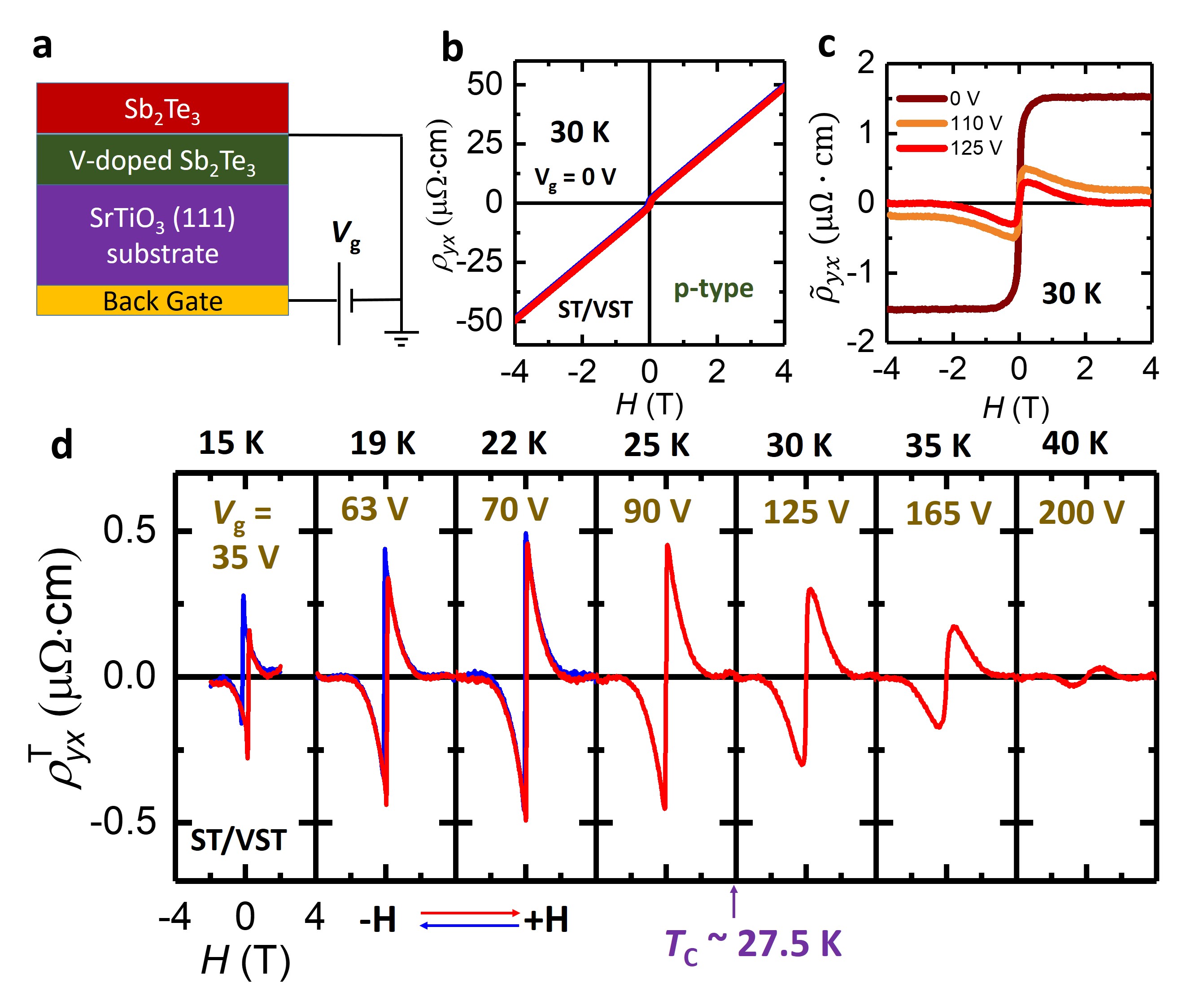}
\caption{\textbf{THE of ST capped VST films} 
\textbf{a}, A schematic of the (ST)$_3$/(VST)$_5$ bilayer grown on STO (111). 
\textbf{b}, $\rho_{yx}$ of ST/VST film at 30\,K. The positive slope at high field indicates $p$-type carriers.
\textbf{c}, $\tilde\rho_{yx}$ of VST film at 30\,K with different gate voltages. The AHE is tuned to zero at 125~V so that only THE is visible.  
\textbf{d}, $\rho^\mathrm{T}_{yx}$ of VST film as a function of magnetic field $H$ at various temperature from 15\,K to 40\,K.  The overall behavior is {the same} as that of SRO films.
\label{fig4}}
\end{figure}
The $H$ dependence of the THE can be understood in the following physical picture.  At the limit of $H\rightarrow0$, the induced $Q$ is proportional to $H$ because each triad carries a net moment as shown in Fig.~\ref{fig3}\textbf{b}. Therefore, $Q\propto H\propto M$ in the paramagnetic phase at $H\rightarrow0$.  At the high field limit  ($H\rightarrow\infty$), the Zeeman energy competes with the DMI, thus suppressing the solid angles $\Omega$ of all triads, resulting in a vanishing $Q$ (and THE). The leading order  in the high field expansion is proportional to $D^2$ because $Q$ respects the spatial inversion symmetry~\cite{Hou2017}, so  $Q\propto(D/H)^2$ as demonstrated by the cyan line in Fig.~\ref{fig3}\textbf{b}. Thus, the overall $H$ dependence can be described by a phenomenological function that interpolate these two limits: 
\begin{equation}
Q(H)\propto\frac{M(H)}{1+(aH)^2}
\label{e:qp}
\end{equation}
where $a\propto 1/D_\mathrm{eff}$ and $D_\mathrm{eff}$ is the effective DMI. This function reasonably describes both MC simulations and experimental data, supporting the simple intuitive picture (See Supplementary information {Fig.\,S18} for detailed analysis).  As shown in Fig.~\ref{fig3}\textbf{c}, the width of THE peak decreases systematically as $t$ increases,  indicating increasing $D_\mathrm{eff}$. This behavior is also corroborated by our MC simulations with varying DMI. (See Supplementary information {Fig.\,S20})  Using Eq.\,(\ref{e:qp}), we extracted the values of $D_\mathrm{eff}\approx1/a$ (up to a numerical factor) of the STO capped (3, 4, 5, and 6~u.c.) and uncapped (6~u.c.) SRO films. $D_\mathrm{eff}\cdot t$ values are shown in Fig.~\ref{fig3}\textbf{c}. Essentially  $D_\mathrm{eff}$ of the STO capped films is inversely proportional to $t$, in excellent agreement with the interface origin of $D_\mathrm{eff}$.  Furthermore, the $D_\mathrm{eff}$ of uncapped SRO film (6\,u.c.) is about twice larger, which is consistent with the stronger inversion symmetry breaking of uncapped SRO film.   The estimated interface DMI value is $D\approx0.2$~meV, which is $1\sim2$ orders of magnitude smaller than that of SrIrO$_3$/SRO interface~\cite{matsuno16}, in good agreement with the expectation of weak DMI in our SRO films. 

To demonstrate the universal nature of spin chirality fluctuation, we present Hall data of 5 quintuples (QL) VST thin film capped with 3 QLs Sb$_2$Te$_3$ (ST).  Because the carrier density of our ST/VST film  is approximately $1000\times$ smaller than that of SRO,  the Fermi level of the ST/VST film can be shifted substantially by a gate voltage. More interestingly,  the AHE can be tuned to zero at every $T$, which allows an unambiguous isolation of the THE signal.  The schematic of the heterostructure is shown in Fig.~\ref{fig4}\textbf{a}. Fig.~\ref{fig4}\textbf{b} shows the raw data ($\rho_{yx}$) of the ST/VST film. The positive slope at high field indicates $p$-type carriers.  An example of gate tuning AHE at 30~K  is shown in Fig.~\ref{fig4}\textbf{c} (See Supplementary information {Fig.~S17} for the complete dataset).  Fig.~\ref{fig4}\textbf{d} shows the THE data at $T$'s above and below $T_\textrm{c}\approx 27.5$~K, showing qualitatively the same $T$-$H$ dependences as that of SRO films.  Furthermore, the sign of THE is consistent with that of carrier type, in excellent agreement with the same  effective magnetic field due to the same spin chirality fluctuations in these two very different 2D ferromagnets with perpendicular anisotropy. 

These excellent agreements unambiguously demonstrate that the THE observed in SRO and VST films originates from the same spin chirality fluctuation in 2D ferromagnets with perpendicular anisotropy.  For $T>T_\mathrm{c}$, the suppression of the THE signal approximately follows a power law, in reasonable agreement with that of MC simulations (see supplementary information {section M}).  This would inspire future studies of the critical scaling behavior of the spin chirality fluctuation. Note that although both systems' ground states are simple ferromagnets with perpendicular anisotropy which is achiral (because $D\ll D_\mathrm{c}$), the thermal fluctuations around $T_\mathrm{c}$ are chiral due to the presence of (weak) chiral interaction $D$.  How $D$ influences the critical behavior of 2D ferromagnets with perpendicular anisotropy is an interesting subject to explore in future studies.  

Although the THE diminishes at $T\ll T_\mathrm{c}$ in STO capped 6~u.c.\ SRO film, the THE persists to low temperatures in the 3~u.c.\ SRO film and the ST/VST film. (See Supplementary information {Fig.\,S11 and S17}).  
It is unclear whether the low temperature THE originates from static topological spin texture (\textit{e.g.}, skrymions), which will inspire future studies of the THE in the Berry phase engineered magnetic thin films such as the ST/VST films.  Our discovery of universal spin chirality fluctuation opens door to explore topological Hall transport in other itinerant 2D ferromagnets and beyond~\cite{Bonilla2018,Fei2018,Deng2018}, and may help to identify chiral spin liquid states with quantum entanglements in 2D correlated systems~\cite{wen89,machida10}.

\vskip 1cm
\noindent {\bf Method}

\noindent {\bf Sample growth and characterization --}
The both the SrRuO$_3$ (SRO) and SrTiO$_3$ (STO) layers were epitaxially grown on TiO$_2$ terminated (001) STO substrates using pulsed laser deposition (PLD) technique. The single TiO$_2$ terminated STO substrates were achieved by standard buffer HF etching for 30\,s and subsequent annealing at 950\,$^\circ$C for 90\,min. During the growth of either SRO or STO, the laser fluence and repetition rate are 2\,J/cm$^2$ and 1\,Hz, respectively. The substrate temperature and oxygen partial pressure during growth are 650$^\circ$C and 0.25\,mBar, respectively. In-situ reflection high-energy electron diffraction (RHEED) was used to monitor the growth and confirmed a layer by layer growth fashion. All the films show atomic flat surface with similar terrace structure with substrates. X-ray diffraction (XRD) has been done by PANalytical X’Pert Materials Research Diffractometer (MRD) in high resolution mode. 

The {Sb$_2$Te$_3$/Sb$_{2-x}$V$_x$Te$_3$ ($x=0.1$)} heterostructure was grown on SrTiO$_3$(111) substrate with molecular beam epitaxy (MBE) method. The insulating SrTiO$_3$ (111) substrates used for the growth was first soaked in 90\,$^\circ$C deionized water for 1.5\,hours, and then annealed at 985\,$^\circ$C for 3\,hours in a tube furnace with flowing pure oxygen gas. Through the above heat treatment, the STO (111) substrate surface become passivated and atomically flat, the topological insulator (TI) heterostructure growth was carried out using a commercial EPI-620 MBE system with a vacuum that is better than $2 \times 10^{-10}$\,mbar. The heat-treated insulating STO (111) substrates were outgassed at $\sim$530\,$^\circ$C for 1\,hour before the growth of the TI heterostructures. High-purity Sb (99.999\%) and Te (99.999\%) were evaporated from Knudsen effusion cells, and V (99.995\%) was evaporated from an e-gun. During growth of the TI, the substrate was maintained at ~240\,$^\circ$C. The flux ratio of Te per Sb was set to be $>$10 to prevent Te deficiency in the samples. The pure or magnetic TI growth rate was at $\sim$0.25 QL/min. Following the growth, the TI films were annealed at $\sim$240\,$^\circ$C for 30\,minutes to improve the crystal quality before being cooled down to room temperature. Finally, to avoid possible contamination, an 18\,nm thick Te layer is deposited at room temperature on top of the sandwich heterostructures prior to their removal from the MBE chamber for transport measurements.

\noindent {\bf Transport measurements --}
The Hall resistance and longitudinal resistance were measured by standard lock-in techniques with an alternating current of 40\,$\mu$A modulated at 314\,Hz. {All Hall data are antisymmetrized.}

\noindent {\bf Monte Carlo simulations -}
The Monte Carlo (MC) simulations were conducted on a $N^2 = 32\times32$ square lattice
with periodic boundary conditions. The Hamiltonian was used
as the energy function underlying the standard Metropolis algorithm. 
The expression of $\Omega$ follows: 
\begin{equation}
\exp\left(\frac{i\Omega}{2}\right)=r^{-1}(1+\bm{S}_1\cdot\bm{S}_2+ \bm{S}_2\cdot\bm{S}_3+ \bm{S}_3\cdot\bm{S}_1+i\chi)
\end{equation}
where $\chi \equiv \bm{S}_1\cdot(\bm{S}_2\times\bm{S}_3)$ is the spin chirality of the three neighboring spins~\cite{wen89}, and $r=\sqrt{2(1+ \bm{S}_1\cdot\bm{S}_2)(1+ \bm{S}_2\cdot\bm{S}_3)(1+ \bm{S}_3\cdot\bm{S}_1)}$ is the normalization factor.  The sign of chirality is fixed by chiral interaction (DMI). The total topological charge (TC) $Q$ is then given by summing $\Omega$ over all the right-handed spin triangles in the lattice,
\begin{equation}
Q = \frac{1}{4\pi}\sum_{\triangle_{ijk}} \Omega(\bm{S}_i,\bm{S}_j,\bm{S}_k).
\end{equation}
$Q$ (or $\chi$) is a odd function with respect to time reversal symmetry, so $Q(H)=-Q(-H)$.  Thus, for paramagnetic phase with $H=0$, the net topological charge is zero, $Q(0)=0$, corresponding to equal populations of ``up'' and ``down'' triads. 

To illustrate chiral fluctuation, we present representative MC simulation results with parameters $D=0.25\,J$ and $K = 0.5\,J$. The simulation of $32\times32$ sites shows a ferromagnetic ordering with $T_\mathrm{c} \approx 0.85\,J$. (The extrapolated thermodynamic $T_\mathrm{c}\approx 0.71\,J$. At $T=J>T_\mathrm{c}$, the $M(H)$ curve shows the expected paramagnetic behavior without hysteresis.

\vskip 0.5cm
\noindent {\bf Code availability --}The MC simulation codes that support the findings of this study are available from the D.X.\ upon reasonable request.

\vskip 0.5cm
\noindent {\bf Data availability --}The data that support the findings of this study are available from the corresponding author upon reasonable request.


\vskip 0.2cm
\noindent {\bf Acknowledgments}\\
We are grateful to Robert Swendsen for many useful discussions on Monte Carlo simulations.  The work at Rutgers is supported by the Office of Basic Energy Sciences, Division of Materials Sciences and Engineering, US Department of Energy under Award numbers DE-SC0018153. D.X. is supported by the Defense Advanced Research Project Agency (DARPA) program on Topological Excitations in Electronics (TEE) under grant number D18AP00011. The work at UMD is supported under the Cooperative Research Agreement between the University of Maryland and the National Institute of Standards and Technology Center for Nanoscale Science and Technology, Award 70NANB14H209, through the University of Maryland.  The work at U. Twente was supported by Nederlandse Organisatie voor Wetenschappelijk Onderzoek through Grant No.13HTSM01.  The work at Penn State is supported by ARO Young Investigator Program Award No.\ W911NF1810198. C.Z.C.\ acknowledges support of Alfred P. Sloan Research Fellowship. 

\vskip 0.2cm
\noindent {\bf Author Contributions}\\
W.Wu conceived and supervised the project. Z.L., J.W., G.K., and G.R.\ synthesized the SRO samples and performed X-ray diffraction. Y.Z.\ and C.-Z.C.\ synthesized the VST samples. W.Wa performed the magnetotransport experiments and analyzed the data. M.D.\ and D.X.\ performed the Monte Carlo simulations. W.Wa, W.Wu, M.D.\ and D.X.\ wrote the manuscript. All authors discussed the data and contributed to the manuscript.

\noindent\textbf{$^\ast$}W.Wa., M.D., and Z.L.\ contributed equally to this work.
\vskip 0.2cm

\noindent {\bf Competing Interests}\\
The authors declare no competing interests.
\vskip 0.2cm


\end{document}